\documentclass[aps,prd,showpacs,preprintnumbers,amsmath,amssymb,nofootinbib,11pt]{revtex4}

\usepackage{tabularx}
\usepackage{here}
\usepackage{graphicx}
\usepackage{subfigure}
\usepackage{epsfig}
\usepackage{color}
\usepackage{slashed}
\usepackage{hyperref}

\def\nin{\noindent}
\def\bea{\begin{eqnarray}}
\def\eea{\end{eqnarray}}

\begin{document}
\pagestyle{plain}

\title{Possible $\Sigma^*(\frac{1}{2}^-)$ in the initial-state polarized $\gamma
N\rightarrow K^{+} \Sigma^*(1385) \rightarrow K^{+} \pi \Lambda$
reaction near threshold}
\author{Yun-Hua Chen$^1$}
\author{ B.~S. Zou$^{1,2}$ }
\affiliation{ ${}^1$ Institute of High Energy Physics, CAS, P.O. Box
918(4), Beijing 100049, China.
\\ ${}^2$ {   State Key Laboratory of Theoretical Physics, Institute of
Theoretical Physics, CAS, Beijing 100190, China.} }

\begin{abstract}
By using an effective Lagrangian method, we study the effects of a
newly proposed  $\Sigma^*(\frac{1}{2}^-)$ state with mass around
1380 MeV in the initial-state polarized $\gamma N\rightarrow K^{+}
\Sigma^*(1385) \rightarrow K^{+} \pi \Lambda$ process near
threshold. The theoretical predictions for the helicity cross
sections $\sigma_{\frac{3}{2}}$, $\sigma_{\frac{1}{2}}$ as well as
their ratios, and the angular distributions of $\pi$ in the
$\pi\Lambda$ center-of-mass system are given. It is found that
assuming $\Sigma^*(\frac{1}{2}^-)$ exists or not, these physical
quantities are distinctly different. So our results could be useful
for the investigation of the existence of $\Sigma^*(\frac{1}{2}^-)$
when the experimental data are available in the future.
\end{abstract}


 \maketitle

\section{Introduction}
\label{intro}

From studies of baryon spectroscopy and internal structures, the
picture of some baryons having large five-quark $qqqq\bar q$
fraction was proposed ~\cite{Helminen,Zou1,Liu,An1,An2}. The
penta-quark picture can naturally solve some puzzles in classic
three-constituent-quark models, for example for the
$J^P=\frac{1}{2}^-$ baryons why $N^*(1535)$ is heavier than
$\Lambda^*(1405)$~\cite{Zou1}. For the lowest mass strange baryon,
the penta-quark models~\cite{Helminen,Zhu} predict a
$\Sigma^{\ast}({1\over2}^-)$ state with mass about 1360$\sim$1405
MeV which is around the mass, 1385 MeV, of the known
$\Sigma^{\ast}({3\over2}^+)$. The studies of $\Sigma^{\ast}$ are of
intrinsic interest to check the correctness of penta-quark models,
and recently some evidence for the existence of the
$\Sigma^{\ast}({1\over2}^-)$ near 1380 MeV has been found through
research on the $K^-p\to \Lambda\pi^+\pi^-$ process~\cite{Wu09,Wu10}
and the
 $K\Lambda\pi$~\cite{Gao10} and $K\Sigma\pi$~\cite{Moriya} photoproduction processes.

Photoproduction of $K\Sigma^{\ast}$ provides a useful tool for
understanding baryon spectroscopy and structures. In the early time
the limited experimental data on the cross section for $\gamma
+p\rightarrow K^++\Sigma^{*0}(1385)$ have large error
bars~\cite{Crouch,Erbe1,Erbe2}. Only in recent years, the
high-statistical experimental data on the $K\Sigma^{\ast}$
photoproduction have been made available. The CLAS Collaboration has
measured the cross section of $\gamma +p\rightarrow
K^++\Sigma^{*0}(1385)$ with photon energies covering from the
threshold up to 4.0 GeV~\cite{CLAS}. The LEPS Collaboration has
reported the first measurement of the cross section and beam
asymmetries of the $\gamma +n\rightarrow K^++\Sigma^{*-}(1385)$
process, using a linearly polarized photon beam with energy of
$E_\gamma=1.5-2.4$~GeV~\cite{LEPS}. Theoretical investigations of
$K\Sigma^{\ast}$ photo-production have been presented in
Refs.~\cite{Lutz,Doring,Nakayama08,Gao10}. In
Ref.~\cite{Nakayama08}, the t-, s-, and u-channel diagrams as well
as the contact term, which are required by gauge invariance, are
calculated and are compared with the CLAS data~\cite{CLAS}. Though
Ref.~\cite{Nakayama08}'s theoretical results of the $K\Sigma^{\ast}$
photoproduction cross section agree well with the CLAS data and LEPS
data, its prediction for the beam asymmetries greatly deviates from
the measurement by the LEPS Collaboration. This obstacle can be
solved by including a new $\Sigma^{\ast}({1\over2}^-)$ state with a
mass around 1380 MeV, and in this way the experimental data from
both the CLAS Collaboration and LEPS Collaboration can be well
described as found in Ref.~\cite{Gao10}.

The existence of $\Sigma^{\ast}({1\over2}^-)$ can also be tested
through the experimental measurement of the initial-state polarized
$\gamma N\rightarrow K^{+} \Sigma^* \rightarrow K^{+} \pi \Lambda$
process. With the photon circularly polarized and the target of the
nucleon polarized along the photon momentum direction, the total
helicity may be $3\over2$ or $1\over2$, corresponding to the
spin-parallel and spin-antiparallel state of the photon and nucleon,
respectively. In the energy range near threshold, the state of total
helicity $3\over2$ can only produce $\Sigma^{\ast}({3\over2}^+)$,
while the the state of total helicity $1\over2$ can produce both
$\Sigma^{\ast}({3\over2}^+)$ and $\Sigma^{\ast}({1\over2}^-)$.
Theoretically, we can predict the helicity cross section
$\sigma_{3\over2}$, $\sigma_{1\over2}$ and the angular distribution
of the final $\pi$ in the $\pi\Lambda$ center-of-mass (c.m.) system
assuming there only exist $\Sigma^{\ast}({3\over2}^+)$ or there
exist both $\Sigma^{\ast}({3\over2}^+)$ and
$\Sigma^{\ast}({1\over2}^-)$. The ratio of
$\frac{\sigma_{3\over2}}{\sigma_{1\over2}}$ and the angular
distribution of $\pi$ will be different in the two cases, so the
existence of $\Sigma^{\ast}({1\over2}^-)$ can be tested by future
experimental analyses. In this article, within the framework of the
gauge-invariant effective Lagrangian from~\cite{Nakayama08,Gao10},
we have made such calculation of the initial-state polarized $\gamma
N\rightarrow K^{+} \Sigma^* \rightarrow K^{+} \pi \Lambda$ process
taking into account or neglecting the $\Sigma^{\ast}({1\over2}^-)$.

This paper is organized as follows. In Sec.~\ref{theor}, the
theoretical framework is presented for the initial-state polarized
$\gamma N\rightarrow K^{+} \Sigma^* \rightarrow K^{+} \pi \Lambda$
process, where $\Sigma^*$ include $\Sigma^*({3\over2}^+)$ and
$\Sigma^*({1\over2}^-)$. In Sec.~\ref{disc}, the theoretical
predictions for the helicity cross sections $\sigma_{3\over2}$,
$\sigma_{1\over2}$, as well as their ratio, and the angular
distribution of the $\pi$ in the $\pi\Lambda$ c.m. system with or
without the $\Sigma^{\ast}({1\over2}^-)$ are presented. We compare
and discuss the results of these two cases. In Sec.~\ref{summary},
we give a summary of this work.

 \nin

\section{Theoretical framework}
\label{theor} \nin

\begin{figure*}[ht]
\begin{center}
\includegraphics[width=0.8\textwidth]{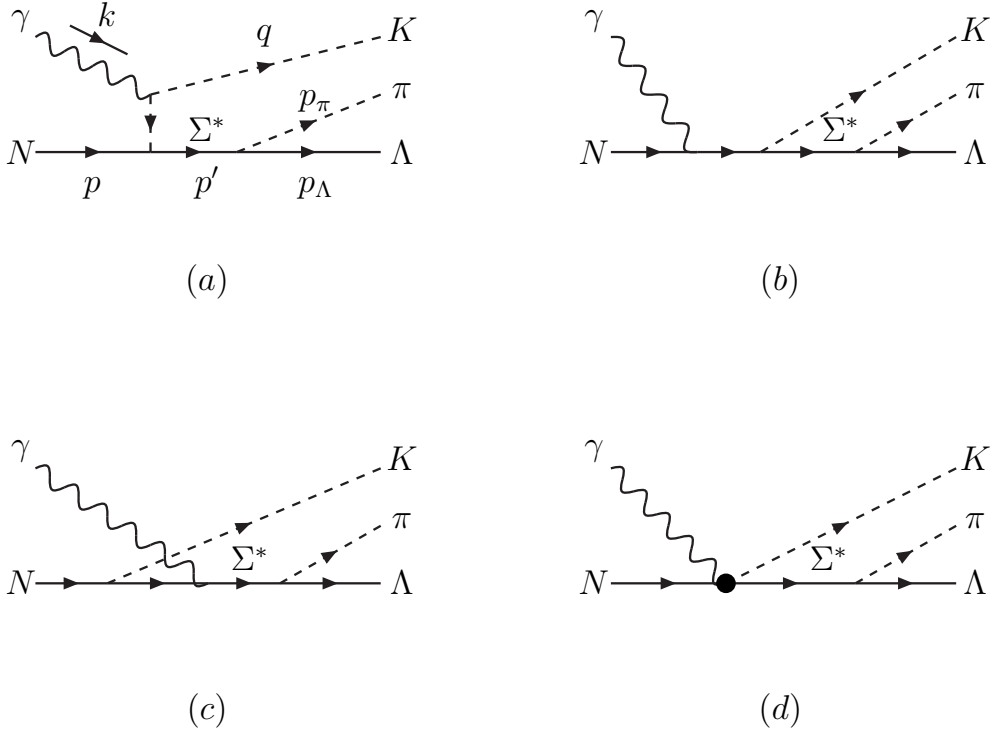}
\caption[pilf]{ Feynman diagrams for $\gamma N\rightarrow K^{+}
\Sigma^* \rightarrow K^{+} \pi \Lambda$. (a) t-channel; (b)
s-channel; (c) u-channel; (d) contact term.}\label{Feynmandiag}
\end{center}
\end{figure*}

The Feynman diagrams for $\gamma N\rightarrow K^{+} \Sigma^*
\rightarrow K^{+} \pi \Lambda$ are shown in Fig.~\ref{Feynmandiag},
where $k$, $p$, $q$, $p_\pi$, and $p_\Lambda$ are the momenta of the
incoming photon and nucleon and outgoing $K$, $\pi$, and $\Lambda$,
respectively, and $p'$ is the momentum of the intermediate
$\Sigma^\ast$. Following the strategy of
Refs.~\cite{Gao10,Nakayama08}, for the reaction $\gamma N\rightarrow
K^{+} \Sigma^*({3\over2}^+) \rightarrow K^{+} \pi \Lambda$ we
consider the contribution of the t-channel $K$ meson exchange, the
s-channel $N$ and $\Delta$ as well as their resonances exchange, the
u-channel $\Lambda$ (for the neutral propagator only) and
$\Sigma^*({3\over2}^+)$ exchange, and the contact term. For the
reaction $\gamma N\rightarrow K^{+} \Sigma^*({1\over2}^-)
\rightarrow K^{+} \pi \Lambda$, we consider the contribution of the
t-channel $K$ meson exchange, the s-channel $N$ exchange, the
u-channel $\Sigma^*({1\over2}^-)$ exchange (and $\Lambda$ exchange
for $\gamma p\rightarrow K^+ \Sigma^{*0}({1\over2}^-)$), and the
contact term.

The effective Lagrangians and coupling constants relevant to the
$\gamma N\rightarrow K^{+} \Sigma^* $ reaction used in this article
are taken from Refs.~\cite{Gao10,Nakayama08} and are listed below
for completeness, and the interested reader can consult
Refs.~\cite{Gao10,Nakayama08} for more details.

For the t-channel $K$ meson exchange:
\begin{equation}\label{gkk}
{\cal L}_{\gamma KK}=i e A_\mu(K^-\partial^\mu K^+-\partial^\mu K^-
K^+)\,,
\end{equation}
\begin{equation}\label{knss}
{\cal L}_{KN\Sigma^*_{3/2}}={f_{KN\Sigma^*_{3/2}}\over
m_K}\partial_\mu\bar{K}{\bar{\Sigma}^{*\mu}_{3/2}}\cdot
{\bf\tau}N+\mathrm{H.c.}\,,
\end{equation}
\begin{equation}\label{kns}
{\cal L}_{KN\Sigma^*_{1/2}}=-i
g_{KN\Sigma^*_{1/2}}\bar{K}{\bar{\Sigma}^*_{1/2}}\cdot
{\bf\tau}N+\mathrm{H.c.}\,,
\end{equation}
with the isospin structure of $K\Sigma^*N$ coupling,
\begin{equation}\label{isostr}
\bar K=(K^-,{\bar K}^0), {\bar{\Sigma}^*}\cdot
{\bf\tau}=\left ( \begin{array}{cc} \bar{\Sigma}^{*0}& \sqrt{2}{\bar\Sigma}^{*+} \\
\sqrt{2}{\bar\Sigma}^{*-} & -{\bar\Sigma}^{*0}
\end{array}\right ),N=\left(\begin{array}{c} p
\\n\end{array}\right),
\end{equation}
where the coupling constant
$f_{KN\Sigma^*_{3/2}}=-3.22\pm0.04$~\cite{Nakayama08} and
$g_{KN\Sigma^*_{1/2}}=1.34\pm0.07$~\cite{Gao10}.

For the s-channel of nucleon exchange, the effective Lagrangian for
the $\gamma NN$ vertex is
\begin{equation}\label{gnn}
{\cal L}_{\gamma NN}=-e \overline{N}(\gamma^\mu A_\mu
Q_N-{\kappa_N\over 2M_N}\sigma^{\mu\nu}\partial_\nu A_\mu)N\,,
\end{equation}
where $Q_N$ is the electric charge (in units of $e$), and $\kappa_N$
 denotes the magnetic moment of the nucleon: $\kappa_n=-1.913$ and $\kappa_p=2.793$.

The $\gamma N\rightarrow K^{+} \Sigma^*({3\over2}^+) $ process has
s-channel spin-${3\over 2}$ and spin-${5\over 2}$ resonances
exchange diagrams, and the effective Lagrangians are
\begin{eqnarray}
{\cal L}_{\gamma NR}({3\over 2}^\pm)&=&-{ief_1\over 2M_N}{\bar
N}\Gamma_\nu^{(\pm)}F^{\mu\nu}R_\mu -{ef_2\over
(2M_N)^2}\partial_\nu{\bar
N}\Gamma^{(\pm)}F^{\mu\nu}R_\mu + \mathrm{H.c.},\\
{\cal L}_{\gamma NR}({5\over 2}^\pm)&=&{ef_1\over (2M_N)^2}{\bar
N}\Gamma_\nu^{(\mp)}\partial^\alpha F^{\mu\nu}R_{\mu\alpha}
-{ief_2\over (2M_N)^3}\partial_\nu{\bar
N}\Gamma^{(\mp)}\partial^\alpha F^{\mu\nu}R_{\mu\alpha} +
\mathrm{H.c.},
\end{eqnarray}
and
\begin{eqnarray}
{\cal L}_{RK\Sigma^*}({3\over 2}^\pm)&=&{h_1\over
m_K}\partial^\alpha
K{\bar\Sigma}^{*\mu}\Gamma_\alpha^{(\pm)}R_\mu+{ih_2\over
(m_K)^2}\partial^\mu\partial^\alpha
K{\bar\Sigma}^*_\alpha\Gamma^{(\pm)}R_\mu+ \mathrm{H.c.},\\
{\cal L}_{RK\Sigma^*}({5\over 2}^\pm)&=&{ih_1\over
m_K^2}\partial^\mu\partial^\beta
K{\bar\Sigma}^{*\alpha}\Gamma_\mu^{(\mp)}R_{\alpha\beta}-{h_2\over
(m_K)^3}\partial^\mu\partial^\alpha\partial^\beta
K{\bar\Sigma}^*_\mu\Gamma^{(\mp)}R_{\alpha\beta}+ \mathrm{H.c.},
\end{eqnarray}
where $F^{\mu\nu}=\partial^\mu A^\nu-\partial^\nu A^\mu$, $R_\mu$
and $R_{\mu\alpha}$ denote the spin-${3\over 2}$ and spin-${5\over
2}$ fields, respectively, and
\begin{equation}
\Gamma_\mu^{(\pm)}=\left(\begin{array}{c} \gamma_\mu\gamma_5
\\\gamma_\mu\end{array}\right), \Gamma^{(\pm)}=\left(\begin{array}{c}
\gamma_5
\\1\end{array}\right).
\end{equation}
For the $\Delta$ resonances of isospin-$\frac{3}{2}$, the effective
Lagrangians have the isospin structure
\begin{eqnarray}
\bar{K} \bar{{\Sigma}}^*\cdot {\bf T}({1\over 2},{3\over 2})
\Delta=\sqrt{3}K^-\bar{\Sigma}^{*+}\Delta^{++}
-\sqrt{2}K^-\bar{\Sigma}^{*0}\Delta^{+}
-K^-\bar{\Sigma}^{*-}\Delta^{0}\nonumber\\
+\bar{K}^0\bar{\Sigma}^{*+}\Delta^{+}
-\sqrt{2}\bar{K}^0\bar{\Sigma}^{*0}\Delta^{0}
-\sqrt{3}\bar{K}^0\bar{\Sigma}^{*-}\Delta^{-}.
\end{eqnarray}

We consider three two-star-rated resonances in the s channel,
$N_{{3\over 2}^-}(2120)$, $\Delta_{{3\over 2}^-}(1940)$, and
$\Delta_{{5\over 2}^+}(2000)$, which are the most prominent
resonances as stated in Ref.~\cite{Nakayama08}. The coupling
constants $f_1$ and $f_2$ can be either computed by using Eq. (B3)
in Ref.~\cite{Nakayama08} from the helicity amplitudes in the
PDG~\cite{PDG} or from the model predictions.
 For the $\gamma N\Delta$ coupling,
we have $f_1=4.04\pm0.20$ and $f_2=3.87\pm0.19$~\cite{Nakayama08}.
From the predicted helicity amplitudes in Ref.~\cite{Capstick92},
one has $f_1=-1.25$ and $f_2=1.21$ for the $\gamma pN^*(2120)$
coupling; $f_1=0.381$ and $f_2=-0.256$ for the $\gamma nN^*(2120)$
coupling; $f_1=0.39$ and $f_2=-0.57$ for the $\gamma N\Delta(1940)$
coupling, and $f_1=-0.68$, $f_2=-0.062$ for the $\gamma N
\Delta(2000)$ coupling~\cite{Gao10}. For the $\Delta K\Sigma^*$
coupling, $h_1=2.000\pm0.006$ and $h_2=0$ are obtained from
$h_1=-f_{K\Delta\Sigma^*}/{\sqrt{3}}$ with
$f_{K\Delta\Sigma^*}=-3.46\pm0.01$~\cite{Nakayama06}. For the
resonances coupling to the $K\Sigma^*$, the coupling constants $h_1$
and $h_2$ can be computed by using Eqs. (B11)-(B18) in
Ref.~\cite{Capstick92} from the model-predicted amplitudes
$G(l)$~\cite{Capstick98}. One obtains $h_1=0.24$ and $h_2=-0.54$ for
the $N^*(2120)K\Sigma^*$ coupling, $h_1=-0.68$ and $h_2=1.0$ for the
$\Delta(1940)K\Sigma^*$ coupling, and $h_1=-1.1$ and $h_2=0.21$ for
the $\Delta(2000)K\Sigma^*$ coupling~\cite{Gao10}. Note that the
masses, widths and coupling constants of the s-channel resonances
$N_{{3\over 2}^-}(2120)$, $\Delta_{{3\over 2}^-}(1940)$, and
$\Delta_{{5\over 2}^+}(2000)$ are not well constrained by the
experiment---hence these parameters have large uncertainties---while
near threshold these three resonances' contributions are very small
so their uncertainties to our theoretical predictions are
negligible.

For the u-channel $\Lambda(1116)$ exchange in the $\gamma
p\rightarrow K^+\Sigma^{*0}$ reaction, the effective Lagrangians are
\begin{eqnarray}
{\cal L}_{\gamma\Lambda\Sigma^*_{3/2}}&=&-{ief_1\over
2M_\Lambda}{\bar
\Lambda}\gamma_\nu\gamma_5F^{\mu\nu}\Sigma^*_{3/2\mu}-{ef_2\over
(2M_\Lambda)^2}\partial_\nu{\bar
\Lambda}\gamma_5F^{\mu\nu}\Sigma^*_{3/2\mu} + \mathrm{H.c.},
\\
{\cal L}_{\gamma
\Lambda\Sigma^*_{1/2}}&=&{eg_{\gamma\Lambda\Sigma^*_{1/2}}\over
4(M_\Lambda+M_{\Sigma^*_{1/2}})}{\bar\Sigma^*_{1/2}}\gamma_5\sigma_{\mu\nu}\Lambda
F^{\nu\mu}+\mathrm{H.c.}\,,
\\
{\cal L}_{KN\Lambda}&=&{g_{KN\Lambda}\over
M_N+M_\Lambda}\bar{N}\gamma^\mu\gamma_5\Lambda\partial_\mu
K+\mathrm{H.c.},
\end{eqnarray}
where $f_1=4.52\pm0.32$, $f_2=5.63\pm0.45$ are obtained from the
decay width $\Gamma(\Sigma^*_{3/2}\rightarrow\Lambda\gamma)$ and
$g_{\gamma\Lambda\Sigma^*_{1/2}}=1.16$. From the flavor SU(3)
symmetry relation, one has
$g_{KN\Lambda}=-13.24\pm1.06$~\cite{Nakayama08}.

For the u-channel $\Sigma^*$ exchange, the effective Lagrangians are
\begin{eqnarray}
{\cal L}_{\gamma \Sigma^*_{1/2}\Sigma^*_{1/2}}&=&-e
\overline{\Sigma}^*_{1/2}(\gamma^\mu A_\mu Q_{\Sigma^*_{1/2}}
-{\kappa_{\Sigma^*_{1/2}}\over 2M_N}\sigma^{\mu\nu}\partial_\nu
A_\mu)\Sigma^*_{1/2}\,,\\
 {\cal
L}_{\gamma\Sigma^*_{3/2}\Sigma^*_{3/2}}&=&e\bar{\Sigma}^*_{3/2\mu}
A_\alpha
\Gamma^{\alpha,\mu\nu}_{\gamma\Sigma^*_{3/2}}\Sigma^*_{3/2\nu},
\end{eqnarray}
with
\begin{eqnarray}
A_\alpha
\Gamma^{\alpha,\mu\nu}_{\gamma\Sigma^*_{3/2}}&=&Q_{\Sigma^*_{3/2}}A_\alpha\big
(g^{\mu\nu}\gamma^\alpha-{1\over
2}(\gamma^\mu\gamma^\nu\gamma^\alpha+\gamma^\alpha\gamma^\mu\gamma^\nu)\big
)-{\kappa_{\Sigma^*_{3/2}}\over
2M_N}\sigma^{\alpha\beta}\partial_\beta A_\alpha g^{\mu\nu},
\end{eqnarray}
where $Q_{\Sigma^*}$ is the electric charge (in units of $e$), and
 $\kappa_{\Sigma^*}$ denotes the anomalous magnetic moment of
 $\Sigma^*$: $\kappa_{\Sigma^{*0}_{3/2}}=0.36$ and
$\kappa_{\Sigma^{*-}_{3/2}}=-2.43$ are taken from the quark model
~\cite{Lich}, and $\kappa_{\Sigma^{*0}_{1/2}}=-0.43$ and
$\kappa_{\Sigma^{*-}_{1/2}}=-1.74$ are predicted by the penta-quark
model~\cite{Zhu}.

To take account of the off-shell effects, every vertex of these
channels has been given a form factor.  For the t-channel $K$ meson
exchange, we use the form factor~\cite{Nakayama08}
\begin{equation}\label{FM}
F_M={\Lambda_M^2-m_K^2\over \Lambda_M^2-q_t^2}\, ,
\end{equation}
where $q_t=k-q$. We adopt $\Lambda_M=0.83$ GeV for $\Sigma^*_{3/2}$
and $\Lambda_M=1.6$ GeV for $\Sigma^*_{1/2}$~\cite{Gao10}. For the
s-channel $N$ and $\Delta$ exchange, the u-channel processes, and
the $\Sigma^* \Lambda\pi$ vertex, we adopt the form
factor~\cite{Nakayama08}
\begin{equation}\label{FB}
F_B(q_{ex}^2,M_{ex})={\Lambda_B^4\over
\Lambda_B^4+(q_{ex}^2-M_{ex}^2)^2}\, ,
\end{equation}
where the $q_{ex}$ and $M_{ex}$ are the 4-momentum and the mass of
the exchanged hadron, respectively. For the s-channel resonances
exchange, the form factor is
\begin{equation}\label{FBR}
F_B(q_s^2,M_R)=\exp\Big(-{(q_s^2-M_R^2)^2\over \Lambda_B^4}\Big)\,.
\end{equation}
with the cutoff parameter $\Lambda_B=1.0$ GeV~\cite{Nakayama08}.
Note in this paper that we only study the near-threshold physics so
the difference between the Gaussian form factors and the more
justifiable dipole form factors is small. We have checked that using
the dipole form factors for all baryons, the numerical differences
are within $1\%$.

The contact term in Fig.~\ref{Feynmandiag}(d) is required to keep
the full amplitude gauge invariant. For the process $\gamma
p\rightarrow K^+\Sigma^{*0}_{3/2}$, we adopt the contact
current~\cite{Nakayama08,Nakayama062}
\begin{equation}
M_c^{\mu\nu}=ie{f_{KN\Sigma^*_{3/2}}\over m_K}(g^{\mu\nu}f_t-q^\mu
C^\nu),
\end{equation}
where $C^\nu$ is expressed as
\begin{eqnarray}\label{cnup}
C^\nu=-(2q-k)^\nu{f_t-1\over t-m_K^2}\big(1-h(1-f_s)\big )
-(2p+k)^\nu{f_s-1\over s-M_N^2}\big(1-h(1-f_t)\big )\,.
\end{eqnarray}
Here the Lorenz indexes $\mu$ and $\nu$ couple to that of
$\Sigma^*_{3/2}$ and the photon, respectively; $f_t=F_M^2$ and
$f_s=F_B^2(s,M_N)$ are form factors squared; and $t=q_t^2$ and
$s=q_s^2$; $h$ is a parameter to be fitted to experiments; and $h=1$
is used in Ref.~\cite{Nakayama08}. For the process $\gamma
p\rightarrow K^+\Sigma^{*0}_{1/2}$, the contact current is
\begin{equation}
M_c^\nu=ie g_{KN\Sigma^*_{1/2}}C^\nu\,,
\end{equation}
where $h=1$ is adopted. For the reaction $\gamma n\rightarrow
K^+\Sigma^{*-}_{3/2}$, the contact current is~\cite{Nakayama062}
\begin{equation}
M_c^{\mu\nu}=ie\sqrt{2}{f_{KN\Sigma^*_{3/2}}\over
m_K}(g^{\mu\nu}f_t-q^\mu C^\nu),
\end{equation}
with
\begin{eqnarray}\label{cnun}
C^\nu=-(2q-k)^\nu{f_t-1\over t-m_K^2}\big(1-h(1-f_u)\big )
+(2p'-k)^\nu{f_u-1\over u-M_{\Sigma^*}^2}\big(1-h(1-f_t)\big ),
\end{eqnarray}
where $f_u=F_B^2(u,M_\Sigma^*)$ is the form factor squared, and
$u=q_u^2$ is the squared momentum transfer for the u channel.
According to Ref.~\cite{Gao10}, $h=1.11$ is taken assuming there
only exist $\Sigma^{\ast}({3\over2}^+)$, and $h=1$ is used if there
exist both $\Sigma^{\ast}({3\over2}^+)$ and
$\Sigma^{\ast}({1\over2}^-)$. For the $\gamma n\rightarrow
K^+\Sigma^{*-}({1\over2}^-)$ process, we adopt the contact current:
\begin{equation}
M_c^\nu=ie\sqrt{2}g_{KN\Sigma^*_{1/2}}C^\nu\,,
\end{equation}
where $C^\nu$ is expressed as Eq. (\ref{cnun}), and here $h=1$ is
taken.

All the ingredients of the $\gamma N\rightarrow K^{+} \Sigma^* $
reaction are given above, and now we list the effective Lagrangians
of the $\Sigma^* \Lambda\pi$ vertex~\cite{Wu10,Gao12}:

\begin{eqnarray}
{\mathcal L}_{ \Lambda \pi\Sigma^{*}_{3/2}}&=&g_{\Lambda
\pi\Sigma^{*}_{3/2}}
\bar{\Lambda}{\Sigma^{*\mu}_{3/2}}\partial_{\mu}\pi+\mathrm{H.c.},\\
{\mathcal L}_{\Lambda\pi\Sigma^{*}_{1/2}}&=&-i
g_{\Lambda\pi\Sigma^{*}_{1/2}}\overline{\Sigma}^{*}_{1/2} \Lambda\pi
+ \mathrm{H.c.},
\end{eqnarray}
where $g_{\Lambda \pi\Sigma^{*}_{3/2}}=9.16\pm0.66$ is obtained from
the decay widths of $\Gamma(\Sigma^*_{3/2}\rightarrow
\Lambda\pi)$~\cite{PDG}, and
$g_{\Lambda\pi\Sigma^{*}_{1/2}}=2.12\pm0.33$ is obtained assuming
the fitted result of the $\Sigma^*_{1/2}$ decay width in
Ref.~\cite{Wu09} is contributed totally by the $\Lambda\pi$ channel.

Further more, we need the propagators of intermediate particles to
calculate the Feymann diagrams. For t-channel exchange $K$ meson,
the propagator is
\begin{equation}
G_{K(q_t)}=1/(q_t^2-m_K^2).
 \end{equation}

For the spin-1/2, spin-3/2 and spin -5/2 baryons the propagators are
respectively
\begin{eqnarray}
G^{\frac{1}{2}}_{R(p)}&=&\frac{\slashed{p}+m}{p^2-m^2},\\
 G^{\frac{3}{2}}_{R(p)}&=&\frac{\slashed{p}+m}{p^2-m^2}\Big(-g^{\mu\nu}+{
\gamma^\mu\gamma^\nu\over 3}+{\gamma^\mu p^\nu -\gamma^\nu
p^\mu\over 3m}+{2p^\mu p^\nu\over 3m^2}\Big),\\
G^{\frac{5}{2}}_{R(p)}&=&\frac{\slashed{p}+m}
{p^2-m^2}S_{\alpha\beta\mu\nu}(p,m),
\end{eqnarray}
where
\begin{eqnarray}
S_{\alpha\beta\mu\nu}(p,m)={1\over 2}({\bar g}_{\alpha\mu}{\bar
g}_{\beta\nu}+{\bar g}_{\alpha\nu}{\bar g}_{\beta\mu})-{1\over
5}{\bar g}_{\alpha\beta}{\bar g}_{\mu\nu}-{1\over
10}({\bar\gamma}_\alpha{\bar\gamma}_\mu{\bar
g}_{\beta\nu}+{\bar\gamma}_\alpha{\bar\gamma}_\nu{\bar
g}_{\beta\mu}+{\bar\gamma}_\beta{\bar\gamma}_\mu{\bar
g}_{\alpha\nu}+{\bar\gamma}_\beta{\bar\gamma}_\nu{\bar
g}_{\alpha\mu}),
\end{eqnarray}
with
\begin{eqnarray}
{\bar g}_{\mu\nu}=g_{\mu\nu}-{p_\mu p_\nu\over m^2},\nonumber\\
{\bar\gamma}_\mu=\gamma_\mu-{p_\mu\over m^2}\slashed{p}.
\end{eqnarray}
For the intermediate resonances with sizable width $\Gamma$, namely
$N_{{3\over 2}^-}(2120)$, $\Delta_{{3\over 2}^-}(1940)$,
$\Delta_{{5\over 2}^+}(2000)$, $\Sigma^{\ast}({3\over2}^+)$, and
$\Sigma^{\ast}({1\over2}^-)$, we replace the denominator $1\over
p^2-m^2$ in the propagators with $1\over p^2-m^2+im\Gamma$, and
replace $m$ in the rest of the propagators with $\sqrt{p^2}$. These
decay widths are taken from Ref.~\cite{Gao10,Wu09}, which are within
the PDG range, $\Gamma_{N^*(2120)}=0.25$ GeV,
$\Gamma_{\Delta(1940)}=0.15$ GeV, $\Gamma_{\Delta(2000)}=0.15$ GeV,
$\Gamma_{\Sigma^*({3\over2}^+)}=0.035\pm0.005$ GeV, and
$\Gamma_{\Sigma^*({1\over2}^-)}=0.119_{-0.035}^{+0.055}$ GeV. Since
previous investigation indicates that the mass of the new
$\Sigma^{\ast}({1\over2}^-)$ is around
$\Sigma^{\ast}({3\over2}^+)$~\cite{Wu09,Wu10,Gao10}, here we assume
its mass be the same as $\Sigma^{\ast}({3\over2}^+)$. Note there are
ambiguities when dealing with the high-spin off-shell
particles~\cite{Pascalutsa,Vrancx,Vereshkov}, since here we are
using a tree-level approach and possible effects might be partially
encoded into the phenomenological coupling constants which are
constrained by the experiments. Also, these uncertainties of
off-shell effects might be partially effectively included into the
form factors, and in this paper the values of the cutoff parameters
$\Lambda_M$ and $\Lambda_B$ are taken from
Refs.~\cite{LEPS,Nakayama08}, gotten by fitting the $\gamma
p\rightarrow K^+\Sigma^{*0}$ data. So the description of high-spin
particles used here can properly explore the phenomenological
physics.

The differential cross section for $\gamma N\rightarrow K^{+}
\Sigma^* \rightarrow K^{+} \pi \Lambda$ can be expressed as
\begin{equation}
d\sigma_{\gamma N\rightarrow K^{+} \Sigma^* \rightarrow K^{+} \pi
\Lambda}={|{\bf q}||{\bf p}_\pi||\bar{\cal{M}}|^2\over (2\pi)^5 32s
|\bf k|} d\Omega d\Omega' d m_{\pi\Lambda}
\end{equation}
where $\bf k$ and $\bf q$ denote the 3-momenta of photon and $K^+$
in the c.m. frame respectively, and ${\bf p}_\pi$ is the 3-momenta
of the produced $\pi$ in the $\Sigma^*$ rest frame; $d\Omega=2\pi
d\cos\theta$, and $\theta$ denotes the angle of the outgoing $K^+$
relative to beam direction in the c.m. frame;
$d\Omega'=d\cos\theta'd\phi'$ is the sphere space of the outgoing
$\pi$ in the $\Sigma^*$ rest frame, and $\theta'$ is the angle
between the $\pi$ direction and the $K^+$ direction in the c.m
system of the $\pi\Lambda$; $m_{\pi\Lambda}$ is the invariant mass
of $\pi$ and $\Lambda$, which satisfies
$m_{\pi\Lambda}^2=(p_\pi+p_\Lambda)^2$. With the z-axis being the
direction of motion of the photon and the x-z plane being the
reaction plane, the polarization vectors for right- and left-handed
photons are
\begin{eqnarray}
\vec{\epsilon}_R=-\frac{1}{\sqrt{2}}(\vec{\epsilon}_x+i\vec{\epsilon}_y),\hspace{0.5cm}
\vec{\epsilon}_L=+\frac{1}{\sqrt{2}}(\vec{\epsilon}_x-i\vec{\epsilon}_y).
\end{eqnarray}
For the polarized nucleon we use the projection operators
~\cite{Bjorken}
\begin{eqnarray}
u(p)\bar{u}(p)=(\slashed{p}+m_N)\frac{1}{2}(1+2\lambda \gamma_5
\slashed{s}),
\end{eqnarray}
where $\lambda=\pm \frac{1}{2}$ is the helicity of the nucleon and
$s=(\frac{|\vec{p}|}{m_N},\frac{E_N}{m_N}\frac{\vec{p}}{|\vec{p}|})$.

\section{Results and discussion}
\label{disc} \nin

\begin{figure}[h]
\begin{minipage}[h]{0.5\textwidth}
\centering \subfigure[ ]{
\includegraphics[width=1.5\textwidth]{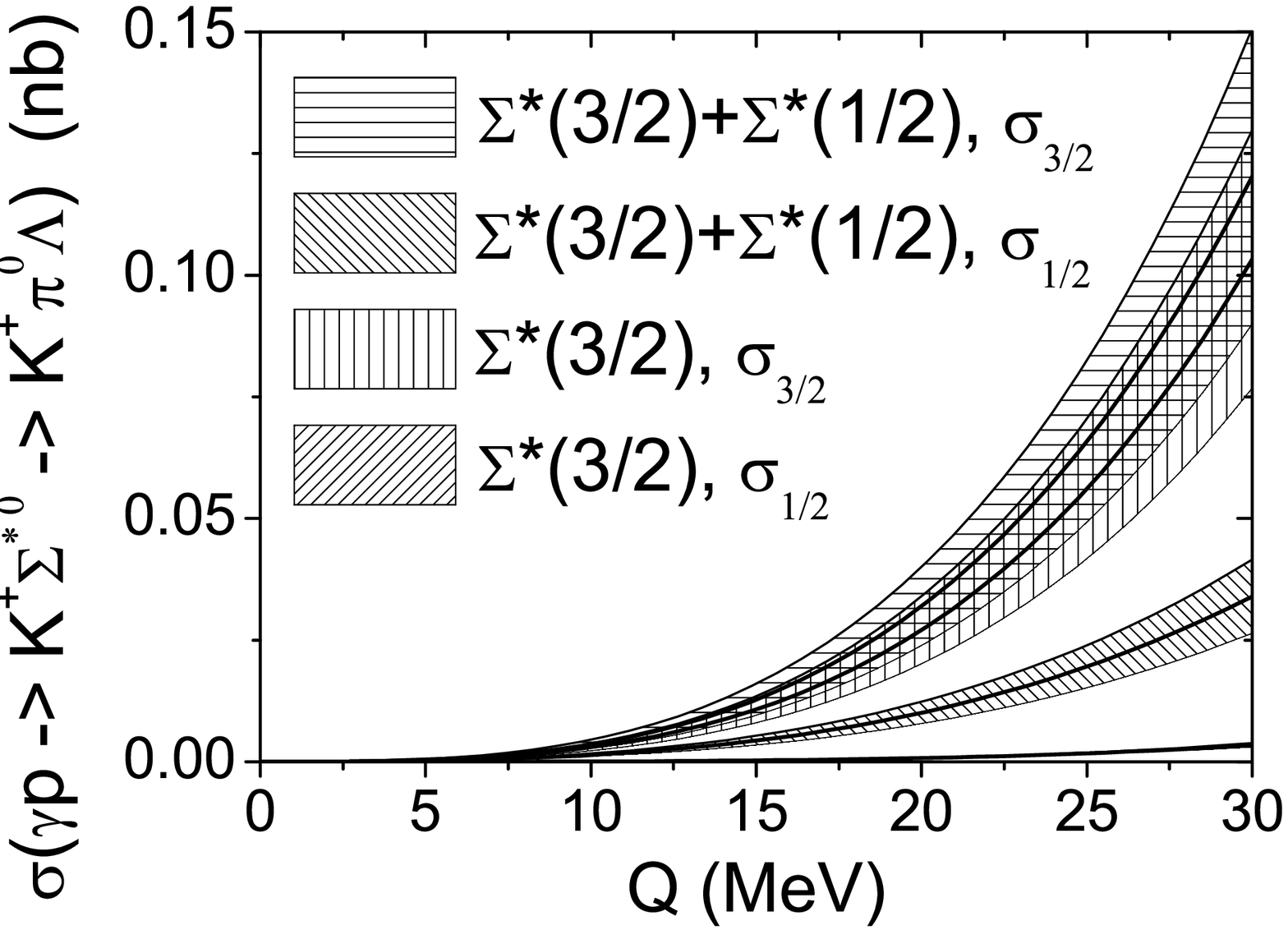}}
\end{minipage}
\begin{minipage}[h]{0.5\textwidth}
\centering \subfigure[]{
\includegraphics[width=1.5\textwidth]{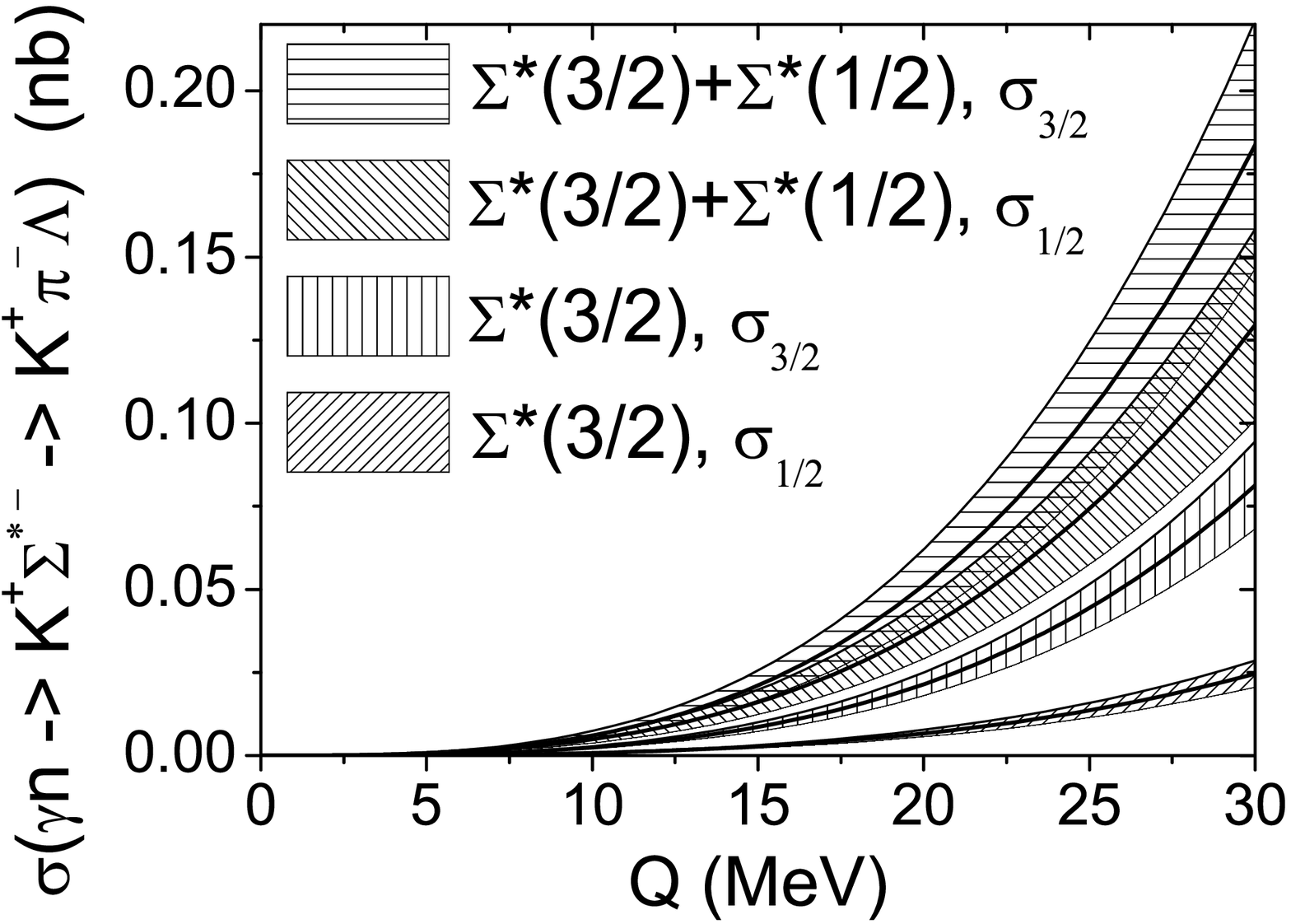}}
\end{minipage}
\caption{ Predictions for the helicity cross sections contributed
from $\Sigma^{\ast}({3\over2}^+)$ and the sum of
$\Sigma^{\ast}({3\over2}^+)$ and $\Sigma^{\ast}({1\over2}^-)$ for
(a) $\gamma p\rightarrow K^{+} \Sigma^{*0} \rightarrow K^{+} \pi^0
\Lambda$ and (b) $\gamma n\rightarrow K^{+} \Sigma^{*-} \rightarrow
K^{+} \pi^- \Lambda$ processes. The shaded areas correspond to the
error bands.}\label{FigTotalCS}
\end{figure}

\begin{figure}[h]
\begin{minipage}[h]{0.5\textwidth}
\centering \subfigure[]{
\includegraphics[width=1.5\textwidth]{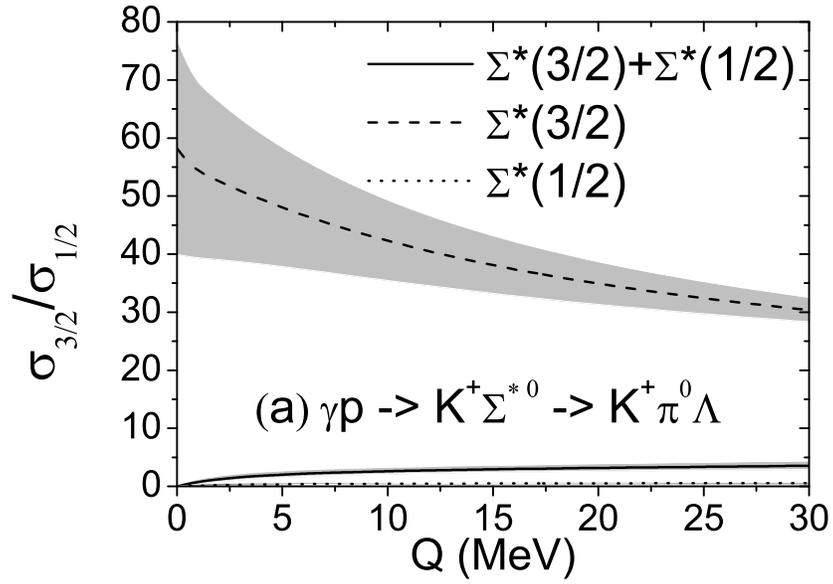}}
\end{minipage}
\begin{minipage}[h]{0.5\textwidth}
\centering
 \subfigure[]{
\includegraphics[width=1.5\textwidth]{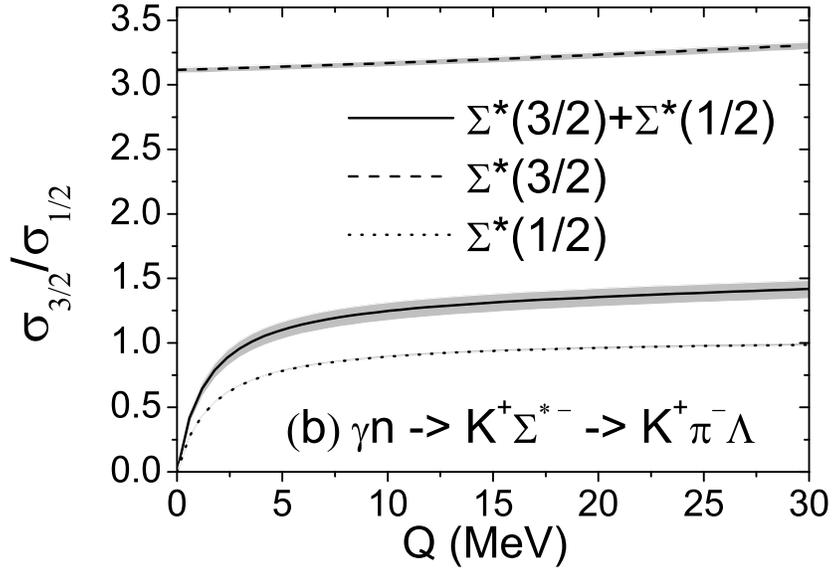}}
\end{minipage}
\caption{ Predictions for the ratios of
$\sigma_{3\over2}/\sigma_{1\over2}$ assuming there exist only
$\Sigma^{\ast}({3\over2}^+)$ (dashed), or only
$\Sigma^{\ast}({1\over2}^-)$ (dotted) or both of them (solid) for
(a) $\gamma p\rightarrow K^{+} \Sigma^{*0} \rightarrow K^{+} \pi^0
\Lambda$ and (b) $\gamma n\rightarrow K^{+} \Sigma^{*-} \rightarrow
K^{+} \pi^- \Lambda$ processes. The shaded areas correspond to the
error bands. }\label{FigCSXoverY}
\end{figure}

With the formalism and ingredients given above, we compute the
helicity cross section $\sigma_{\frac{3}{2}}$ and
$\sigma_{\frac{1}{2}}$, corresponding to spin-parallel and
spin-antiparallel states of the photon and nucleon, respectively,
for the $\gamma N\rightarrow K^{+} \Sigma^* \rightarrow K^{+} \pi
\Lambda$ process assuming there only exists
$\Sigma^{\ast}({3\over2}^+)$ or there exist both
$\Sigma^{\ast}({3\over2}^+)$ and $\Sigma^{\ast}({1\over2}^-)$. The
cross sections versus excess energy in the c.m. frame,
Q$=\sqrt{s}-\sqrt{s}_{threshold}$, are shown in
Fig.~\ref{FigTotalCS}. In Fig.~\ref{FigCSXoverY}, the behavior of
the ratios of $\sigma_{3\over2}/\sigma_{1\over2}$ is given. The
error bands are computed in this way: First we compute the maximum
and the minimum of each theoretical prediction with the coupling
constants within the range of error, then we take
(maximum-minimum)/2 as the error bar of corresponding prediction.

Through analysis we find that the contact terms and the u-channel
$\Lambda$ exchange give the most important contributions to the
$\gamma p\rightarrow K^{+} \Sigma^{*0}(\frac{3}{2}^+) \rightarrow
K^{+} \pi^0 \Lambda$ process, while their interference term enhances
and reduces the total cross section for $\sigma_{\frac{3}{2}}$ and
$\sigma_{\frac{1}{2}}$, respectively, so the ratio of
$\sigma_{3\over2}/\sigma_{1\over2}$ for the pure
$\Sigma^{\ast}({3\over2}^+)$ produced process is about 40 as in
Fig.~\ref{FigCSXoverY} (a). For the $\gamma p\rightarrow K^{+}
\Sigma^{*0}(\frac{1}{2}^-) \rightarrow K^{+} \pi^0 \Lambda$ process,
$\sigma_{\frac{1}{2}}$ comes mainly from the t-channel $K$ exchange
and the s-channel $N$ exchange, while in $\sigma_{\frac{3}{2}}$ the
s-channel $N$ exchange's contribution is suppressed due to angular
momentum conservation so $\sigma_{\frac{1}{2}}$ is larger than
$\sigma_{\frac{3}{2}}$. Assuming there exist both
$\Sigma^{\ast}({3\over2}^+)$ and $\Sigma^{\ast}({1\over2}^-)$, the
ratio of $\sigma_{3\over2}/\sigma_{1\over2}$ is about 3 which is
distinct from that assuming only $\Sigma^{\ast}({3\over2}^+)$ exist,
which can be seen in Fig.~\ref{FigCSXoverY} (a).

For the $\gamma n\rightarrow K^{+} \Sigma^{*-}(\frac{3}{2}^+)
\rightarrow K^{+} \pi^- \Lambda$ process, the contact term plays the
major role and its contribution to the total cross section is two
orders larger than those from other channels, so the ratio of
$\sigma_{3\over2}/\sigma_{1\over2}$ mainly depends on the behavior
of the contact term. For the $\gamma n\rightarrow K^{+}
\Sigma^{*-}(\frac{1}{2}^-) \rightarrow K^{+} \pi^- \Lambda$ process,
the major contribution is from the t-channel $K$ exchange. As can be
seen in Fig.~\ref{FigCSXoverY} (a) and (b), the ratios of
$\sigma_{3\over2}/\sigma_{1\over2}$ from pure
$\Sigma^{\ast}({1\over2}^-)$ are zero at threshold as expected,
while they sharply rise and reach about one when Q$=10$ MeV. This is
because the amplitude of the major t-channel $K$ exchange in
$\Sigma^{\ast}({1\over2}^-)$ produced reactions is proportional to
the component of the photon polarization vector parallel to the
reaction plane, and its contributions to the total cross section are
the same for right and left handed photons. According to our
calculated results, the $\Sigma^{\ast}({1\over2}^-)$ produced cross
sections are larger than those produced by
$\Sigma^{\ast}({3\over2}^+)$, so taking account of the
$\Sigma^{\ast}({1\over2}^-)$ or not, both the total cross section
$\sigma_{\frac{3}{2}}$ and $\sigma_{\frac{1}{2}}$ are different, as
shown in Fig.~\ref{FigTotalCS} (b). Also, in Fig.~\ref{FigCSXoverY}
(b), the ratios of $\sigma_{3\over2}/\sigma_{1\over2}$ are different
assuming there exist both $\Sigma^{\ast}({3\over2}^+)$ and
$\Sigma^{\ast}({1\over2}^-)$ or only exist
$\Sigma^{\ast}({3\over2}^+)$.

\begin{figure*}[ht]
\begin{center}
\includegraphics[width=1.0\textwidth]{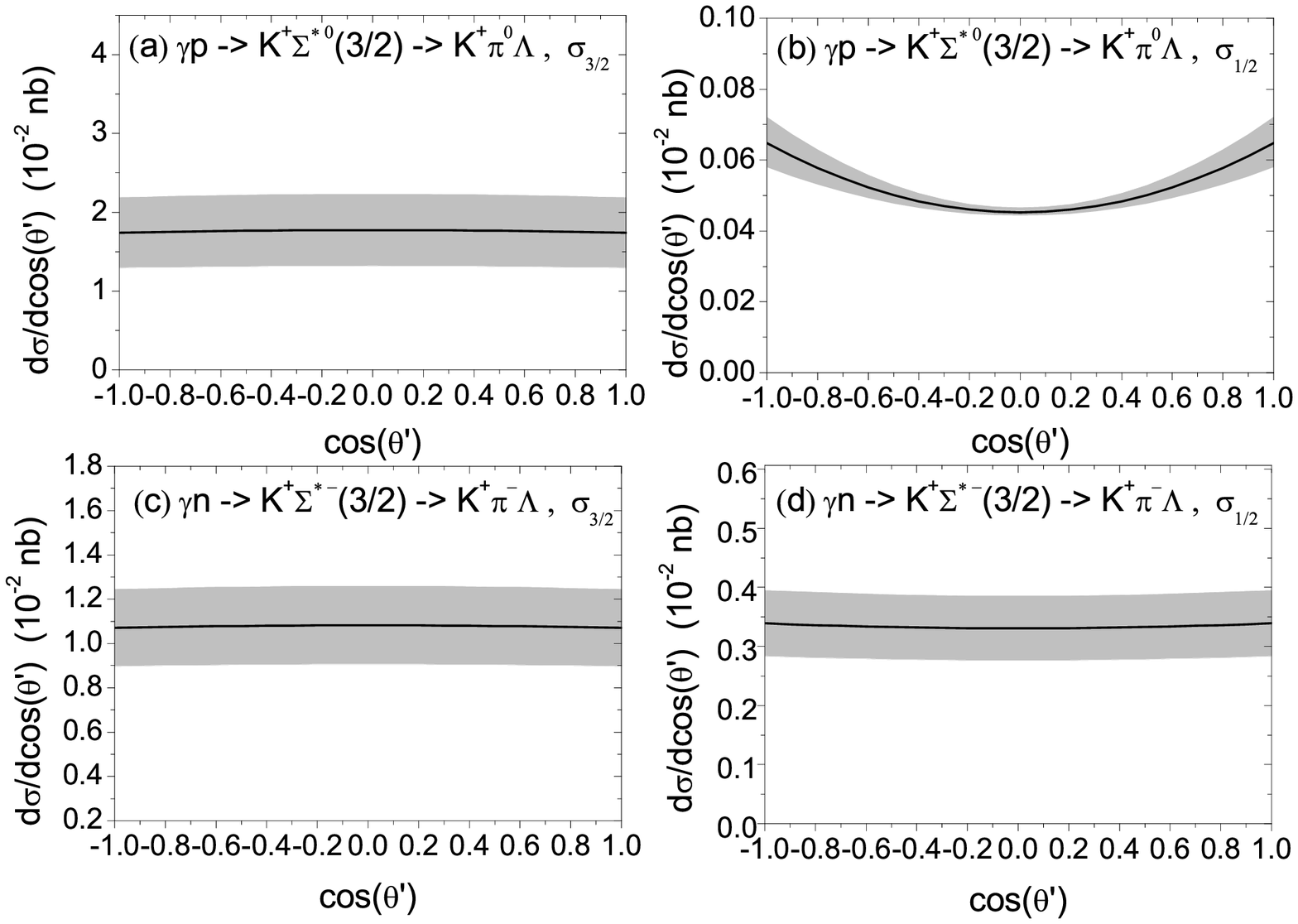}
\caption{ Predictions for the angular distribution of final $\pi$ of
the $\gamma N\rightarrow K^{+} \Sigma^{*}(\frac{3}{2}^+) \rightarrow
K^{+} \pi \Lambda$ process, where $\theta'$ is the angle between the
outgoing $\pi$ direction and $K$ direction in the c.m. system of
$\pi\Lambda$. (a) and (b) denote $\sigma_{3/2}$ and $\sigma_{1/2}$,
respectively, for $\gamma p\rightarrow K^{+}
\Sigma^{*0}(\frac{3}{2}^+) \rightarrow K^{+} \pi^0 \Lambda$ process.
(c) and (d) denote $\sigma_{3/2}$ and $\sigma_{1/2}$, respectively,
for $\gamma n\rightarrow K^{+} \Sigma^{*-}(\frac{3}{2}^+)
\rightarrow K^{+} \pi^- \Lambda$ process. The shaded areas
correspond to the error bands.}\label{FigAngular1Res}
\end{center}
\end{figure*}

\begin{figure*}[ht]
\begin{center}
\includegraphics[width=1.0\textwidth]{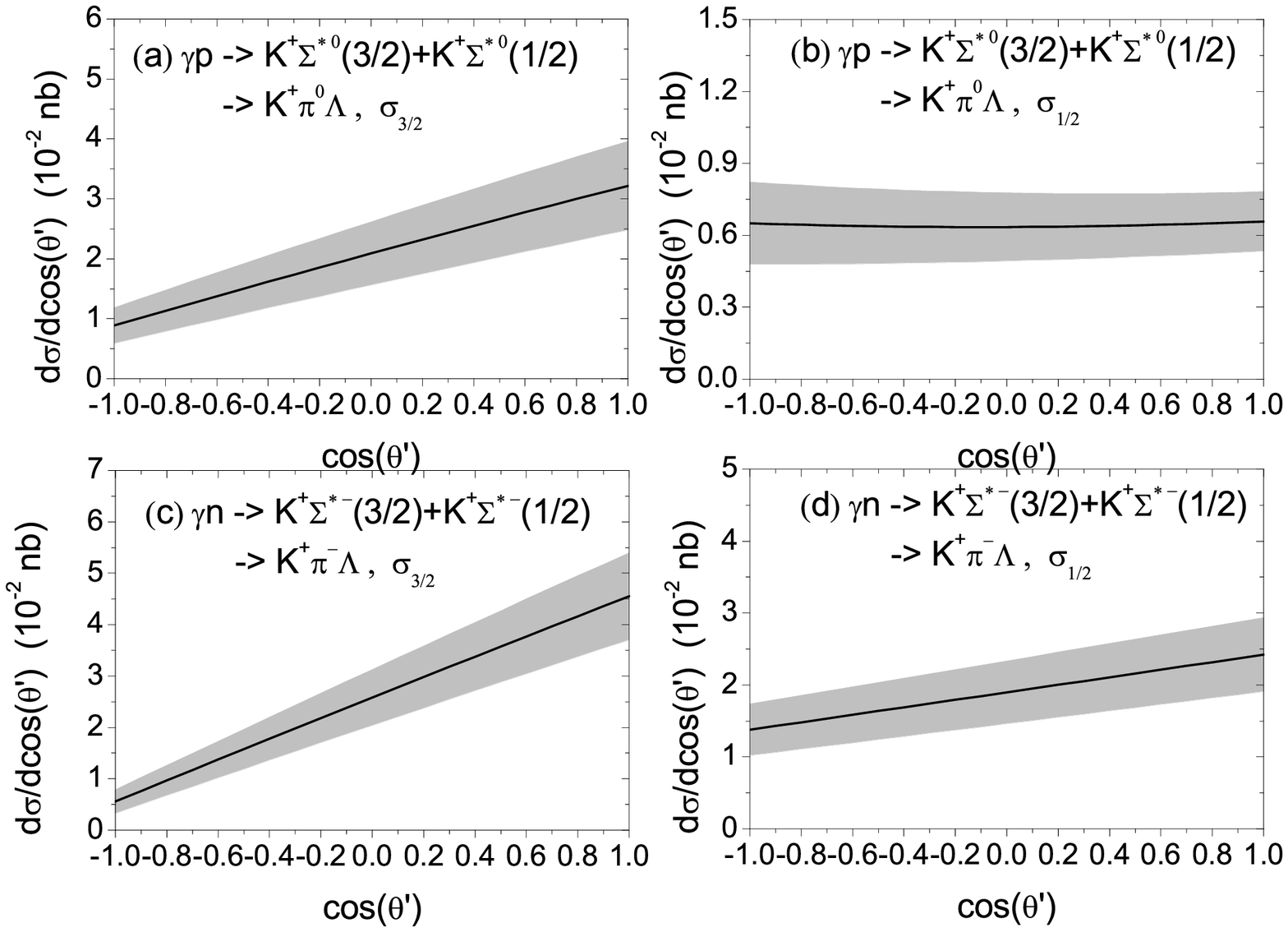}
\caption{Predictions for the angular distribution of final $\pi$ of
the $\gamma N\rightarrow K^{+} \Sigma^{*} \rightarrow K^{+} \pi
\Lambda$ process, where $\Sigma^{*}$ include
$\Sigma^{*}(\frac{3}{2}^+)$ and $\Sigma^{*}(\frac{1}{2}^-)$. (a) and
(b) denote $\sigma_{3/2}$ and $\sigma_{1/2}$, respectively, for
$\gamma p\rightarrow K^{+} \Sigma^{*0} \rightarrow K^{+} \pi^0
\Lambda$ process. (c) and (d) denote $\sigma_{3/2}$ and
$\sigma_{1/2}$, respectively, for $\gamma n\rightarrow K^{+}
\Sigma^{*-} \rightarrow K^{+} \pi^- \Lambda$ process. The shaded
areas correspond to the error bands.}\label{FigAngular2Res}
\end{center}
\end{figure*}

Another way to investigate the spin of the $\Sigma^*$ is to utilize
the angular distribution of the $\pi$ in the $\pi\Lambda$
center-of-mass system. Near threshold, the final $\pi\Lambda$ state
is in the relative $p$ wave from the decay of
$\Sigma^{\ast}({3\over2}^+)$ and is in the relative $s$ wave from
the decay of $\Sigma^{\ast}({1\over2}^-)$. So the angular
distribution is expected to be of the form $(a+b \cos\theta'^2)$ for
the pure $\Sigma^{\ast}({3\over2}^+)$ and a flat constant
distribution is predicted for pure $\Sigma^{\ast}({1\over2}^-)$. In
Fig.~\ref{FigAngular1Res} and Fig.~\ref{FigAngular2Res}, we show the
angular distribution of the $\pi$ in the $\pi\Lambda$ center-of-mass
system for the $\gamma N\rightarrow K^{+} \Sigma^{*} \rightarrow
K^{+} \pi \Lambda$ process assuming there exist only
$\Sigma^{\ast}({3\over2}^+)$ and there exist both
$\Sigma^{\ast}({3\over2}^+)$ and $\Sigma^{\ast}({1\over2}^-)$ at
Q$=20 $ MeV, respectively. Note that here we choose the energy Q$=20
$ MeV just as an example, and the behaviors of the angular
distributions do not change significantly near threshold. As
illustrated in Fig.~\ref{FigAngular1Res}, the shapes of angular
distributions for pure $\Sigma^{\ast}({3\over2}^+)$ agree well with
the expectations. We also have checked that the predictions for the
angular distributions from pure $\Sigma^{\ast}({1\over2}^-)$ are
flat constants, and we do not illustrate them individually in the
figures. The differential cross section contributed by the
interference terms of the $\Sigma^{\ast}({3\over2}^+)$ and
$\Sigma^{\ast}({1\over2}^-)$ are linear functions of $\cos\theta'$,
and we find they change much more rapidly than the corresponding
pure $\Sigma^{\ast}({3\over2}^+)$ terms in the $\gamma p\rightarrow
K^{+} \Sigma^{*0} \rightarrow K^{+} \pi^0 \Lambda$ process for
$\sigma_{3/2}$, and in the $\gamma n\rightarrow K^{+} \Sigma^{*-}
\rightarrow K^{+} \pi^- \Lambda$ process for $\sigma_{3/2}$ and
$\sigma_{1/2}$, so in these reactions the interference terms mainly
determine the shapes of the angular distributions as shown in
Figs.~\ref{FigAngular2Res}(a),~\ref{FigAngular2Res}(c),
and~\ref{FigAngular2Res}(d). In the $\gamma p\rightarrow K^{+}
\Sigma^{*0} \rightarrow K^{+} \pi^0 \Lambda$ process for
$\sigma_{1/2}$, the interference term changes more slowly than the
pure $\Sigma^{\ast}({3\over2}^+)$ term so the shape of the angular
distribution deviates slightly from that of pure
$\Sigma^{\ast}({3\over2}^+)$, as can be seen in
Fig.~\ref{FigAngular2Res}(b).

\newpage

\section{Summary}
\label{summary}

In this paper, we study the reactions $\gamma N\rightarrow K^{+}
\Sigma^{*}(1385) \rightarrow K^{+} \pi \Lambda$ near threshold
within an effective Lagrangian approach. Recent studies indicate
that near the mass of $\Sigma^{*}(\frac{3}{2}^+)$, another
$\Sigma^*$ state with $J^P=\frac{1}{2}^-$ may exits. The spin of
$\Sigma^{*}$ can be investigated in the $K\Sigma^*$ photoproduction
process using circularly polarized photons and a target of polarized
nucleons. Taking account of the $\Sigma^{*}(\frac{1}{2}^-)$ or not,
we compute the helicity cross sections $\sigma_{\frac{3}{2}}$ and
$\sigma_{\frac{1}{2}}$, which correspond to spin-parallel and
spin-antiparallel states of the photon and nucleon respectively, and
their ratios. Also we give the predictions for the angular
distributions of the $\pi$ in the $\pi\Lambda$ c.m. system. Through
the analysis, we find that the $\Sigma^{*}(\frac{1}{2}^-)$ and the
interference term of $\Sigma^{*}(\frac{3}{2}^+)$ and
$\Sigma^{*}(\frac{1}{2}^-)$ play significant roles near threshold,
such that the ratios of $\sigma_{\frac{3}{2}}/\sigma_{\frac{1}{2}}$
and the angular distribution of the $\pi$ are distinctly different
assuming that the $\Sigma^{*}(\frac{1}{2}^-)$ exists or not. The
results of this work may be useful for identification of
$\Sigma^{*}(\frac{1}{2}^-)$ when the experimental data are available
in the future.

\section*{Acknowledgements}

We acknowledge Eulogio Oset for the suggestion to start this work
and for carefully reading through the manuscript. We also thank Puze
Gao, Jia-jun Wu, and Jian-ping Dai for helpful discussions. This
work is supported in part by the National Natural Science Foundation
of China under Grants No. 11035006, No. 11121092, and No.
11261130311 (CRC110 by DFG and NSFC), the Chinese Academy of
Sciences under Project No. KJCX2-EW-N01, and the Ministry of Science
and Technology of China (2009CB825200).


\begin{thebibliography}{99}

\bibitem{Helminen} C.~Helminen and D.~O.~Riska, Nucl.\ Phys.\ A {\bf 699}, 624 (2002).

\bibitem{Zou1} B.~S.~Zou, Eur.\ Phys.\ J.\ A {\bf 35}, 325 (2008); Int.\ J.\ Mod.\ Phys.\ A {\bf 21}, 5552 (2006).

\bibitem{Liu} B.~C.~Liu and B.~S.~Zou, Phys.\ Rev.\ Lett.\ {\bf 96}, 042002
(2006); Phys.\ Rev.\ Lett.\ {\bf 98}, 039102 (2007).

\bibitem{An1}  C.~S.~An, Q.~B.~Li, D.~O.~Riska and B.~S.~Zou,  Phys.\ Rev.\ C {\bf 74},
055205 (2006); Phys.\ Rev.\ C {\bf 75}, 069901(E) (2007).

\bibitem{An2}  C.~S.~An, Nucl.\ Phys.\ A {\bf 797}, 131 (2007); Nucl.\ Phys.\ A {\bf 801}, 82
(2008).

\bibitem{Zhu} A.~Zhang \textit{et al.}, High Energy Phys. Nucl. Phys.\ {\bf 29}, 250 (2005), arXiv:hep-ph/0403210.

\bibitem{Wu09}  J.~J.~Wu, S.~Dulat, and B.~S.~Zou,  Phys.\ Rev.\ D {\bf 80}, 017503
(2009).

\bibitem{Wu10}  J.~J.~Wu, S.~Dulat, and B.~S.~Zou,  Phys.\ Rev.\ C {\bf 81}, 045210 (2010).

\bibitem{Gao10} P.~Z.~Gao, J.~J.~Wu, and B.~S.~Zou,  Phys.\ Rev.\ C {\bf 81},
055203 (2010).

\bibitem{Moriya}
  K.~Moriya {\it et al.}  [CLAS Collaboration],
  arXiv:1301.5000 [nucl-ex].

\bibitem{Crouch} J.~H.~R.~Crouch  \textit{et al.} (Cambridge Bubble Chamber Group), Phys.\ Rev.\ {\bf 156},
1426 (1967).

\bibitem{Erbe1} R.~Erbe \textit{et al.} (DESY Bubble Chamber Group), Nuovo Cimento A {\bf 49},
504 (1967).

\bibitem{Erbe2} R.~Erbe \textit{et al.} (ABBHHM Collaboration), Phys.\ Rev.\ {\bf 188},
2060 (1969).


\bibitem {CLAS} L.~Guo and D.~P.~Weygand (CLAS Collaboration),
 in \textit{Proceedings of the International Workshop on the Physics of
 Excited Baryons (NSTAR05)}, edited by S.~Capstick, V.~Crede, and P.~Eugenio
 (World Scientific, Singapore, 2006, pp. 306-309.

\bibitem{LEPS} K.~Hicks \textit{et al.} (LEPS Collaboration), Phys.\ Rev.\ Lett.\ {\bf 102},
012501 (2009).

\bibitem{Lutz} M.~F.~M.~Lutz and M.~Soyeur, Nucl.\ Phys.\ A {\bf 748}, 499 (2005).

\bibitem{Doring} M.~D\"{o}ring, E.~Oset, and D.~Strottman,, Phys.\ Lett.\ B {\bf 639}, 59
(2006); Phys.\ Rev.\ C {\bf 73}, 045209 (2006).

\bibitem{Nakayama08} Y.~Oh, C.~M.~Ko, and K.~Nakayama,  Phys.\ Rev.\ C {\bf 77},
045204 (2008).

\bibitem{PDG}  J.~Beringer \textit{et al.} (Particle Data Group),  Phys.\ Rev.\ D {\bf 86}, 010001 (2012).

\bibitem{Capstick92}  S.~Capstick,  Phys.\ Rev.\ D {\bf 46}, 2864 (1992).

\bibitem{Nakayama06} Y.~Oh, K.~Nakayama, and T.-S.~H.~Lee, Phys.\ Rep.{\bf 423},
49 (2006).

\bibitem{Capstick98}  S.~Capstick and W.~Roberts,  Phys.\ Rev.\ D {\bf 58}, 074011 (1998).

\bibitem{Lich}  D.~B.~Lichtenberg,  Phys.\ Rev.\ D {\bf 15}, 345 (1977).

\bibitem{Nakayama062} H.~Haberzettl, K.~Nakayama, and S.~Krewald,  Phys.\ Rev.\ C {\bf 74},
045202 (2006).

\bibitem{Gao12} P.~Z.~Gao, J.~Shi, and B.~S.~Zou,  Phys.\ Rev.\ C {\bf 86},
025201 (2012).

\bibitem{Pascalutsa} V.~Pascalutsa, R.~Timmermans,  Phys.\ Rev.\ C {\bf 60},
042201 (1999).

\bibitem{Vrancx} T.~Vrancx, L.~De Cruz, J.~Ryckebusch and P.~Vancraeyveld,  Phys.\ Rev.\ C {\bf 84},
045201 (2011).

\bibitem{Vereshkov} G.~Vereshkov, N.~Volchanskiy, Phys.\ Rev.\ C {\bf 87},
035203 (2013).

\bibitem{Bjorken} J.~D.~Bjorken and S.~D.~Drell, \textit{Relativistic Quantum
Mechanics}, Graw-Hill, New York, 1965.



\end{thebibliography}
\end{document}